\documentclass[preprint,12pt]{elsarticle}

\newcommand{\beginsupplement}{%
        \setcounter{table}{0}
        \renewcommand{\thetable}{S\arabic{table}}%
        \setcounter{figure}{0}
        \renewcommand{\thefigure}{S\arabic{figure}}%
     }

\usepackage{graphicx}
\usepackage{amssymb}
\usepackage{url}
\usepackage[utf8]{inputenc}

\journal{Journal Name}

\begin{document}

\begin{frontmatter}


\title{Local risk perception enhances epidemic control}



\author[ictp,austin]{Jos\'e L. Herrera}
\author[austin]{Lauren Ancel Meyers}
\address[austin]{Department of Integrative Biology, The University of Texas at Austin. Texas - USA}
\address[ictp]{ICTP South American Institute for Fundamental Research
IFT-UNESP, S\~ao Paulo, SP Brazil  01440-070}

\begin{abstract}
As infectious disease outbreaks emerge, public health agencies often enact vaccination and social distancing measures to slow transmission. Their success depends on not only strategies and resources, but also public adherence. Individual willingness to take precautions may be influenced by global factors, such as news media, or local factors, such as infected family members or friends. Here, we compare three modes of epidemiological decision-making in the midst of a growing outbreak. Individuals decide whether to adopt a recommended intervention based on overall disease prevalence, the proportion of social contacts infected, or the number of social contacts infected. While all strategies can substantially mitigate transmission, vaccinating (or self isolating) based on the number of infected acquaintances is expected to achieve the greatest herd immunity and number of infections averted, while requiring the fewest intervention resources.
\end{abstract}




\end{frontmatter}


\section{Introduction}
\label{S:1}
As outbreaks emerge, public health agencies often implement a variety of pharmaceutical and non-pharmaceutical interventions to prevent epidemic expansion, including vaccination and medical prophylaxis, school closures and other social distancing measures, and information campaigns to promote awareness, hygienic precautions and voluntary isolation \cite{serpell,koh,samit,nph}. However, such measures require population adherence and are often hindered by failure to take recommended actions \cite{larson}. Around the globe, for example, seasonal influenza vaccine coverage falls significantly below the 75\% baseline recommended by the World Health Organization, but varies widely between countries and across age groups \cite{levelsVacc}. In the USA, 2015-2016 uptake was only 59.3\% in children and 41.7\% in adults \cite{CDC:estimVacc}.
For measles, routine childhood vaccination is declining in Texas and other areas of the United States where personal belief and other non-medical vaccination exemptions are allowed \cite{hotez,omer,mapVacc}. 
Parental decision-making regarding childhood vaccines is complex and context dependent \cite{larson1}, but likely influenced by false claims regarding vaccine safety, low perceived risks of infectious diseases, and other forms of misinformation from the "anti-vaxxer" movement \cite{hotez,larson1,sage,chapman}. Recently, there have been calls for a special government commission on vaccine safety, despite overwhelming scientific consensus that vaccines are both safe and effective \cite{sage,vaccSafe,jain,taylor1}.

As outbreaks unfold, people can take a variety precautionary measures to avoid infection, including immunization and social distancing \cite{serpell,lau,philipson,ahituv}. They often judge personal risk based on their impressions of overall disease prevalence and severity \cite{koh,funk,durham,zwart}. When infection risk appears low, small risks of adverse affects from the vaccine can seem relatively important and cause vaccine coverage to drop below levels required to control transmission.  
A variety of other factors can influence the perceived utility of disease prevention, including epidemiological news and rumors, costs of vaccination and other control measures, trust in health professionals, government agencies, media and non-official information sources, as well as societal pressure to ensure the health of one's children \cite{thomas,hackett,brown,samba,myers,fabry}. 

Studies have shown that media reports about outbreaks that specify numbers of cases, hospitalizations or deaths, can influence avoidance behavior and contact patterns at both individual and community levels. In some cases, oversimplified media reports regarding flawed vaccines can trigger panics that lead to drops in vaccination and vaccinated mixing with the infectives without regard to disease risk \cite{liu,cui,liY,tchuenche}. For both seasonal and pandemic influenza, such interactions between vaccination decision-making and transmission dynamics can profoundly shape the course of epidemics  \cite{funk,cornforth,feng,andrews}.

Individual intervention decisions can have far-reaching effects. For example, vaccination protects not only the immunized individual, but also social contacts who might have been infected by the individual; social distancing decisions may break chains of transmission by protecting the decision-maker and more generally disrupting social dynamics. The extent of herd immunity (indirect protection) will critically depend on contact patterns \cite{chainBi1,pastor1}. Measures taken by gregarious individuals may have greater immediate benefits than those taken by solitary individuals, with downstream epidemiological consequences modulated by the full social network \cite{pastor,goltsev}. Contact patterns may also influence the decision-making process itself, by constraining epidemiological perspectives. When gauging infection risk, individuals may consider \textit{global} information (e.g., from news media) or first-hand encounters with disease (e.g., infected acquaintances, friends or family members) \cite{liu, cui}. While traditional compartmental models assume homogeneity in both epidemiological risks and intervention benefits, network-based models provide a tractable framework for studying the complex interplay between contact networks, intervention decision making and disease transmission \cite{cornforth,lauren1,ndeffo,herrera}.   

Here, we investigate the epidemiological impacts of different decision paradigms using a network-based SIR epidemic model, in which individuals make vaccination or social distancing choices based on their perceived epidemiological risks. Depending on the decision model, they estimate either overall disease prevalence, their number of infected social contacts, or their fraction of infected social contacts. When the perceived threat is sufficiently high, they take a measure that immediately affords full protection for the duration of the epidemic. We compare the efficacy of these three different paradigms across a range of diseases in a realistically heterogeneous network, and show that the most naive model--simply counting one's infected contacts--affords the most epidemiological protection using the least amount of resources (e.g., vaccinations or economic costs associated with social distancing).

\section{Methods}

We simulate the spread of an infectious disease in a network (population) with an exponential degree distribution--as estimated for typical urban populations \cite{ndeffo,sarsR0}--using a susceptible-infected-recovered (SIR) chain-binomial model \cite{chainBi1,chainBi}. At each time step, individuals decide whether or not to take an instantaneously protective action to avoid infection, based on their perceived risk of infection, as defined by the given local or global decision model. We assume that there are sufficient resources to immediately protect any willing individuals. 

\subsection{Contact network}

We model contact patterns in the population using an exponential network with $N=10000$ nodes and mean number of contacts $\mu=10$ \cite{bansal}, generated according to the configuration model \cite{newman}, unless otherwise is specified. We assume that this network constrains both disease transmission and local risk perceptions, when individuals monitor infected social contacts.
To evaluate the impact of network topology, we also analyze the SIR-intervention dynamics on a homogeneous random graph (all nodes have same degree) and Barab\'asi-Albert scale-free network \cite{barabasi}, with degree distributions constrained to achieve the same epidemic threshold as the focal exponential network. All three networks share $T_c=\frac{\langle k \rangle}{\langle k^2 \rangle - \langle k \rangle}=0.056$   where $\langle k \rangle$ and $\langle k^2 \rangle$ are the average degree and squared degree in the network, respectively.

\subsection{Epidemic dynamics}

We model \emph{SIR} transmission dynamics of a flu-like disease using chain-binomial stochastic  simulations \cite{chainBi1,chainBi}. Epidemics begin at time $t=0$ by infecting a single randomly chosen node in an otherwise completely susceptible population and terminate when there are no remaining infected individuals. Individuals remain infectious for $l=7$ days before recovering with full immunity to future infection \cite{daysFlu,daysFlu1}. Infected individuals transmit disease to each susceptible contact at a rate $\beta$. Immunized and recovered individuals are assumed to be fully resistant to infection.
Results are averages over 500 simulations, unless otherwise specified.   

The basic reproduction number ($R_0$) is the expected number of secondary cases when a single case of disease is introduced into a naive population, and is related to the epidemic growth rate. To study the impact of transmission rate on the vaccination-epidemiological dynamics, we consider $R_0$ values ranging between one and ten, which spans the range for many common human pathogens, including influenza, Ebola, SARS, Pertussis, HIV/AIDS, Mumps, Rubella, Polio, Smallpox, Diphtheria, etc \cite{ebolaR0,ebolaR01,pertussisR0,sarsR0,sarsR01,rsiR0,dpmR0,hivR0,hivR01}. For each value of $R_0$, we determine the corresponding $\beta$ using \cite{lauren1}
\begin{equation}
R_0=T \left( \frac{\langle k^2\rangle - \langle k \rangle}{\langle k \rangle} \right),
\end{equation}
where $T=1-(1-\beta)^l$ is the  transmissibility over the entire infectious period and $\langle k^2\rangle$ is the mean squared degree of nodes in the network. 

\subsection{Immunization models}

\begin{figure}[!ht]
\centering\includegraphics[width=0.9\linewidth]{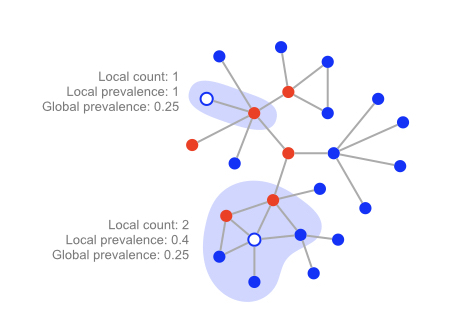} 
\caption{Three decision strategies. Individuals decide to vaccinate based on one of three risk measures: The number of infected contacts (local count), the fraction of infected contacts (local prevalence), or the overall fraction infected (global prevalence). In this example, six of the 24 nodes are infected, yielding a global prevalence of 0.25. The white node towards the top has a single contact that happens to be infected; the white node towards the bottom has two of its five contacts infected.}
\label{strats}
\end{figure}

We assume that individuals make daily decisions regarding whether or not to take precautionary measures based on their perceived risk (Figure \ref{strats}). If and when they choose to take action, they instantly gain full resistance to infection for the remainder of the simulation. Although these models apply to any transmission preventing measure, we henceforth refer to the interventions as \emph{vaccinations}. 

We model three different individual decision strategies in which individuals consider either the disease states of their direct social contacts or the global situation, perhaps gleaned through news or social media. Let $v_i(t)$ denote the willingness of a individual to vaccinate under strategy $i$ at time $t$.

\subsubsection{Local decision strategies}

In the first model, \textit{local prevalence}, individuals assess infection risk by tracking the fraction of their social contacts that are currently infected. The probability that a susceptible individual $i$  vaccinates at time $t$ is given by 
\begin{equation}
v{\textsubscript{lp}}(i,t) = 1-(1-T)^{\frac{\eta_i(t)}{k_i}\times \langle k\rangle} 
\label{four}
\end{equation}
where $\eta_i(t)$ is the number of neighbors of $i$ that are infected at time $t$, $k_i$ is the total number of neighbors (degree) of $i$, and $\langle k \rangle$ is the average degree of the network.  

In the second model, \textit{local count}, individuals track their number rather than proportion of infected neighbors, and decide to vaccinate according to
\begin{equation}
v\textsubscript{lc}(i,t)=1-(1-T)^{\eta_{i(t)}}.
\label{five}
\end{equation}

Local prevalence is arguably a less plausible than local count, given that the decisions require the additional knowledge of total number of contacts (degree) of each individual.

\subsubsection{Global decision strategy}

The \textit{global prevalence model} assumes that individuals base their vaccination decisions on the epidemiological state of the entire population, as given by
\begin{equation}
v\textsubscript{g}(i,t)=1-(1-T)^{\frac{I(t)}{N}\times\langle k \rangle}, \label{six}
\end{equation}
where $I(t)$ is the total number of infected individuals in the population at time $t$ and $N$ is the size of the population. This assumes general knowledge of the evolving dynamics of the epidemic, perhaps through news, social media or public health messaging.

The mean degree ($\langle k \rangle$) appears in the global prevalence and local prevalence as a normalizer. If node $i$ has the average degree ($k_i=\langle k \rangle$) and its local prevalence mirrors global prevalence ($\frac{\eta_i(t)}{\langle k \rangle}=\frac{I(t)}{N}$), then it will have the same probability of vaccinating across all three models. 

In all three models, we assume that individuals will vaccinate with a probability equal to their perceived real-time probability of being infected. For example, if an individual believes that they have a $25\%$ chance of being infected in the immediate future, then they will take precautionary measures with probability 0.25.  The local count model comes closest to estimating actual risk of infection. Specifically, $v\textsubscript{lc}$ is the probability that any currently infected contact will transmit disease to the focal node at some point during his or her infections period. This exactly estimates risk if all infected contacts were just infected and at the beginning of their infectious period, but overestimates risk if some are nearing recovery.

\subsection{Herd immunity}

To assess the indirect and direct protection afforded by a given decision strategy $D$ at a given $R_0$, we calculate a quantity we call \textit{herd immunity}, given by

\begin{equation}
H(D,R_0) =\frac{\langle C_{0}(R_0) \rangle - \langle C_{D}(R_0)\rangle}{\langle V_{D}(R_0)\rangle} ,
\label{herdEq}
\end{equation}
where $\langle C_{0}(R_0) \rangle$ and $\langle C_{D}(R_0)\rangle$ are the expected total number infections in epidemics without vaccination and with vaccination decision strategy $D$, respectively, and $\langle V_{D}(R_0)\rangle$ is the expected total number of individuals vaccinated under $D$. We estimate these expected values by averaging over 500 simulations with the specified $R_0$ and decision model. Barring extreme stochasticity, we expect $H>0$ for any reasonably protective vaccine strategy. When $H$ is between zero and one, more vaccines are given than infections averted, suggesting that vaccines may be mistimed or misplaced. This could happen, for example, if risk is underestimated early in the epidemic and overestimated late in the epidemic. An $H$ near one indicates that approximately one infection is averted for every vaccine given. Note that this is an average, and does not necessarily mean that every vaccination prevents infection of the recipient. If each vaccine averts, on average, greater than one infection ($H>1$), then the value of $H$ corresponds to the level of indirect protection or \textit{herd immunity} achieved by the decision strategy.

\section{Results}

\begin{figure}[!ht]
\centering\includegraphics[width=0.99\linewidth]{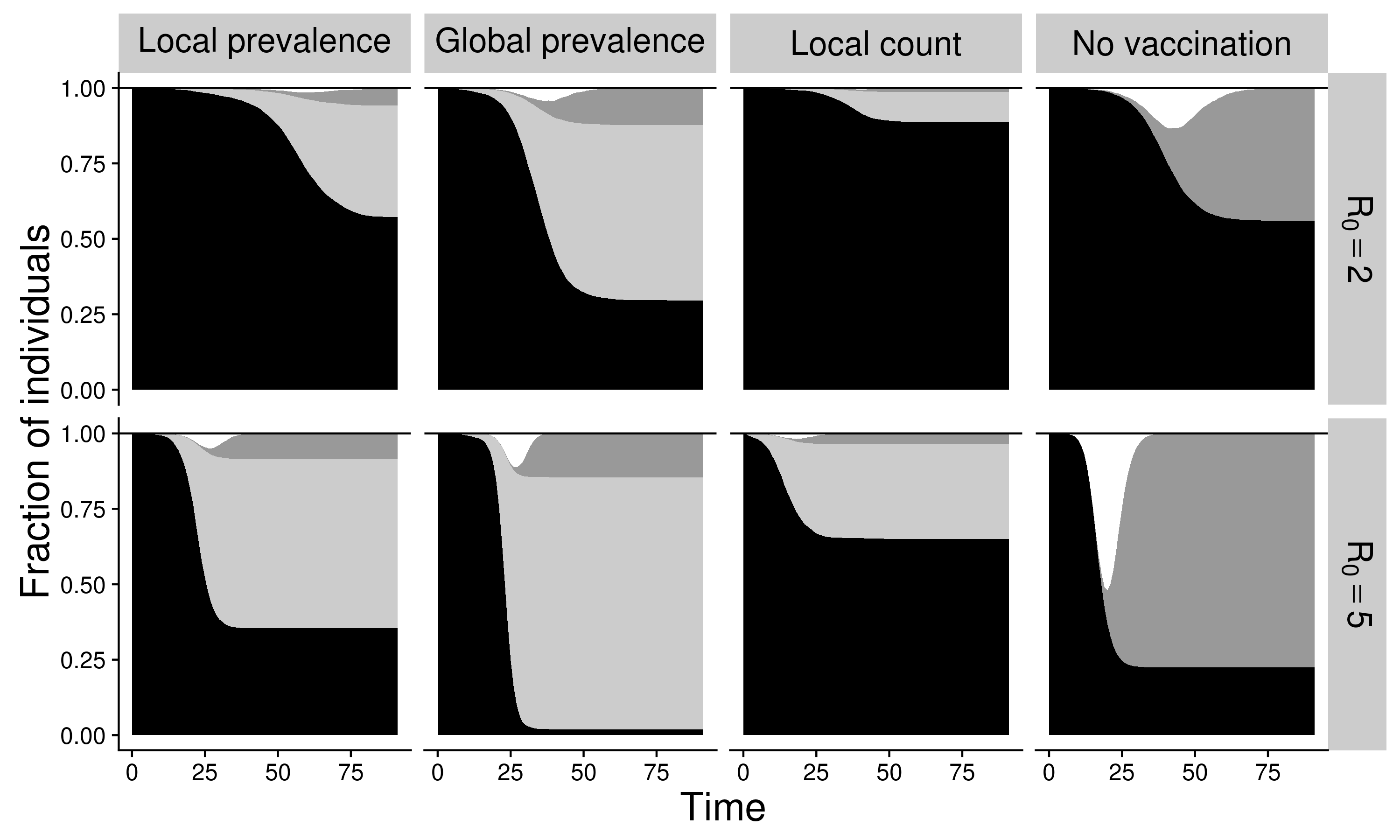}
\caption{Disease and vaccination dynamics under different decision models. Shading indicates the fraction of nodes in each state: susceptible (black), vaccinated (light gray), recovered (dark gray), infected (white). Columns corresponds to different strategies, as labeled above; rows correspond to $R_0=2$ (top) and $R_0=5$ (bottom).}
\label{tempo}
\end{figure}

The decision models yield distinct vaccine adoption and disease transmission dynamics (Figure \ref{tempo}). As disease begins to spread, individuals perceive increasing risks and vaccinate according to the decision model, thereby protecting themselves and interrupting potential chains of transmission to others. While all three strategies reduce the total number of infections, the local count strategy affords the greatest and most efficient protection of the three. Under the global prevalence strategy, perceived risk is homogeneous. As cases mount, the vaccination rate rises synchronously throughout the population, arguably resulting in \textit{too much too late} vaccine coverage. 

\begin{figure}[!ht]
\centering\includegraphics[width=\linewidth]{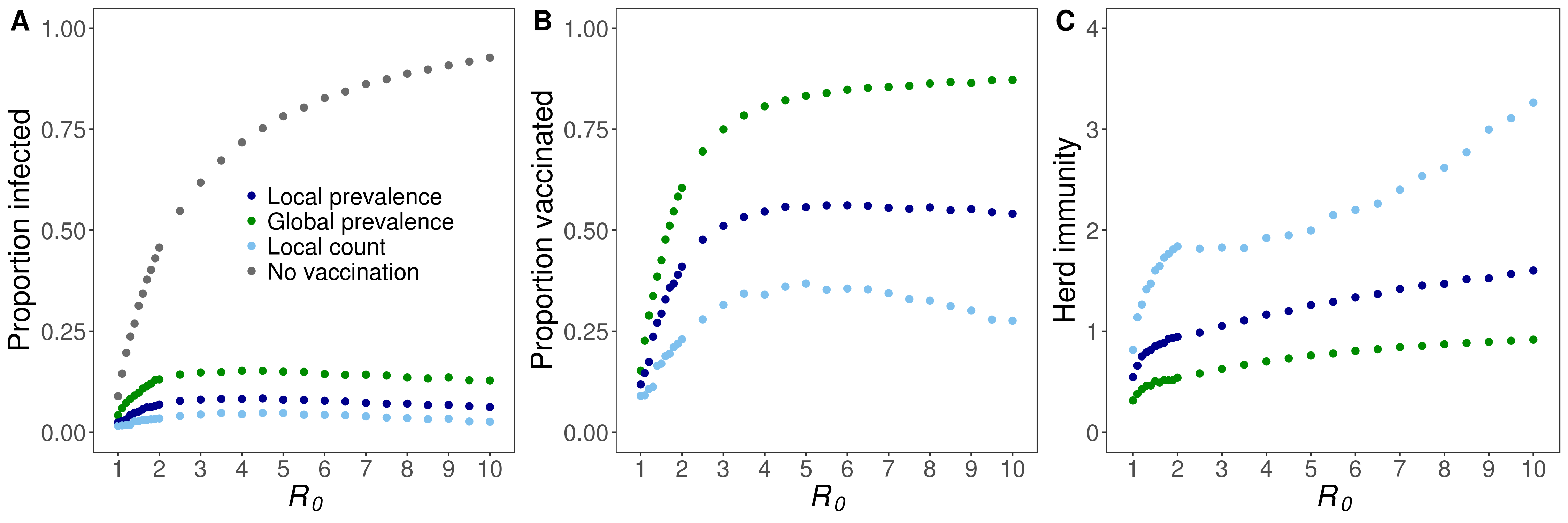}
\caption{Epidemiological impacts of decision strategies change with $R_0$. (A) The proportion of individuals that are infected increases and then declines slightly as $R_0$ increases, across all three decision strategies. (B) The proportion of individuals that choose to vaccinate initially increases sharply with $R_0$ across all three strategies, but then declines for only the {\it local count} strategy. (C) Herd immunity is the proportion of infections averted per vaccinated individual, and is highest for the {\it local count} strategy across all $R_0$. All values are averages across 500 stochastic simulations.
}
\label{asymp}
\end{figure}

The local strategies avert more infections with fewer vaccinations than the global strategy. As epidemics unfold, risk is both heterogeneous and dynamic, with some portions of the network experiencing greater forces of infection than others. Local decision-making allows earlier detection and response to increasing personal risk, and prevents unnecessary vaccination in lower risk settings, both prior to and following epidemic waves. The local count strategy is more protective than the local prevalence strategy. By tracking the number rather than proportion of infected contacts, individuals more accurately assess the local force of infection. For example, compare a solitary individual with just two social contacts and a gregarious individual with $20$. If they both have two infected contacts, then their risk of infection will be similar (assuming that time spent with each contact is sufficient for transmission). Under local count, their perceived risk and consequent vaccination probability will be identical; under local prevalence, the solitary individual will perceive higher risk (i.e., $100\%$ of contacts infected) than the gregarious individual. Under all models, overall vaccine coverage increases as $R_0$ increases, with the global prevalence achieving near universal coverage by $R_0=5$.

\begin{figure}[!ht]
\centering\includegraphics[width=\linewidth]{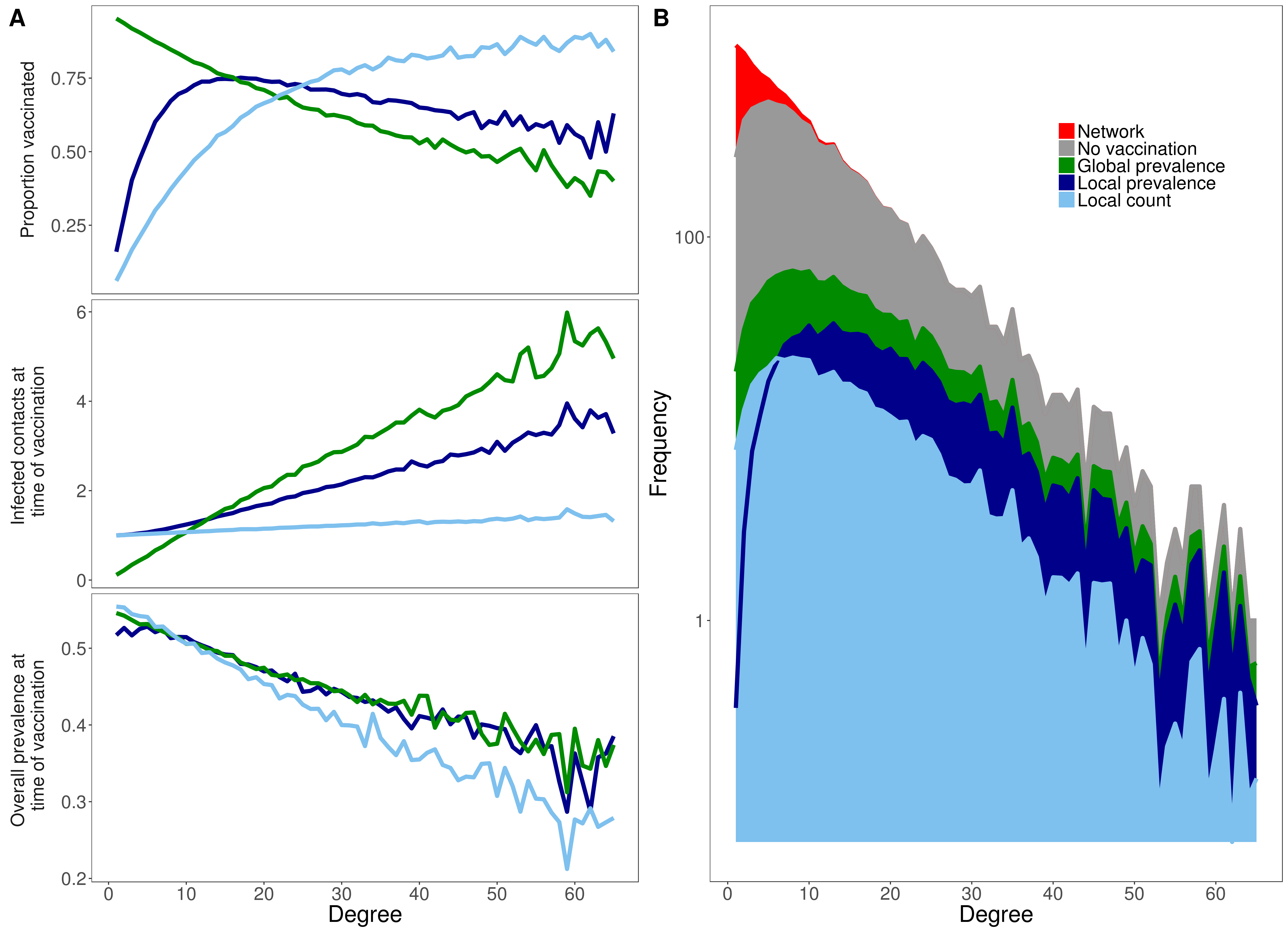}
\caption{Vaccination decisions vary with degree. For each degree class, we graph (A) the proportion of individuals that vaccinate (top) along with the epidemiological situation at the time a node chooses to vaccinate in terms of the number of its infected neighbors (middle) and the overall disease prevalence in the population (bottom). (B) The overall exponential degree distribution in the population (red) on a log scale and the numbers of individuals in each degree class that become infected without vaccination (gray) and under each of the different vaccination strategies. Epidemiological risks---the chances of both becoming infected and infecting others---generally increase with degree. The local count strategy (light blue) is the only strategy for which the probability of vaccinating consistently increases with risk. Compared to the two other strategies, high degree individuals vaccinate earlier in terms of both local and global disease prevalence and, consequently, are less likely to become infected. Values are averages across 500 simulations, assuming $R_0=5$.}
\label{timeVacc1}
\end{figure}

The relative and absolute impacts of each strategy are remarkably robust to the transmissibility of the pathogen (Figure \ref{asymp}). Without vaccines, the expected epidemic size increases non-linearly with $R_0$, reaching almost $100\%$ by $R_0=10$ (Figure \ref{asymp}A). All of the vaccine strategies avert a large and increasing fraction of cases, as $R_0$ increases. In fact, the total epidemic size is non-monotonic, with slightly more expected infections around $R_0=4$ than around $R_0=10$. The local count strategy consistently yields the greatest protection, followed by local prevalence. 

The decision models result in dramatically different vaccination rates, with the global prevalence strategy leading to near universal vaccination, consistently more than double the coverage produced by the local count strategy (Figure \ref{asymp}B). The population-level protection afforded by the local count strategy exhibits a non-trivial trend with $R_0$ (Figure \ref{asymp}C). Between $R_0=1$ and $R_0=2$, its impact grows logarithmically from less than one infection averted per vaccinator to nearly two infections averted per vaccinator. The indirect benefits then appear to grow exponentially, reaching three infections averted per vaccinator when $R_0=9$. 

To explore the dynamic interactions between behavior and epidemiology in the three models, we consider individual nodes based on their degree (number of contacts). In general, the higher the degree of a node, the higher their risk for becoming infected and infecting others, and the greater the number of local infections they could potentially perceive when making vaccination decisions. Indeed, across all decision models, higher degree individuals vaccinate earlier, in terms of the fraction of the population infected at time of vaccination (Figure \ref{timeVacc1}A, bottom). However, when we consider the fraction of individuals vaccinated in each degree class, the three models do \emph{not} consistently afford protection commensurate with risk (Figure \ref{timeVacc1}A, top). Local count is the only strategy that results in vaccination coverage that monotonically increases with risk. Under global decision-making, coverage is inversely related to risk, and under local prevalence, coverage peaks for moderately connected individuals. 
By the time individuals choose to vaccinate under the either of two suboptimal strategies, their local risk of infection is already quite high (Figure \ref{timeVacc1}A, middle), particularly for more gregarious individuals. Although the vaccinating individuals are immediately protected, comparable individuals (of the same degree class and local risk) who stochastically fail to make the same low probability vaccination decision are likely to become infected. Consequently, the risk of infection increases steeply with degree under all models except local count (Figure \ref{timeVacc1}B).

\begin{figure}[!ht]
\centering\includegraphics[width=0.9\linewidth]{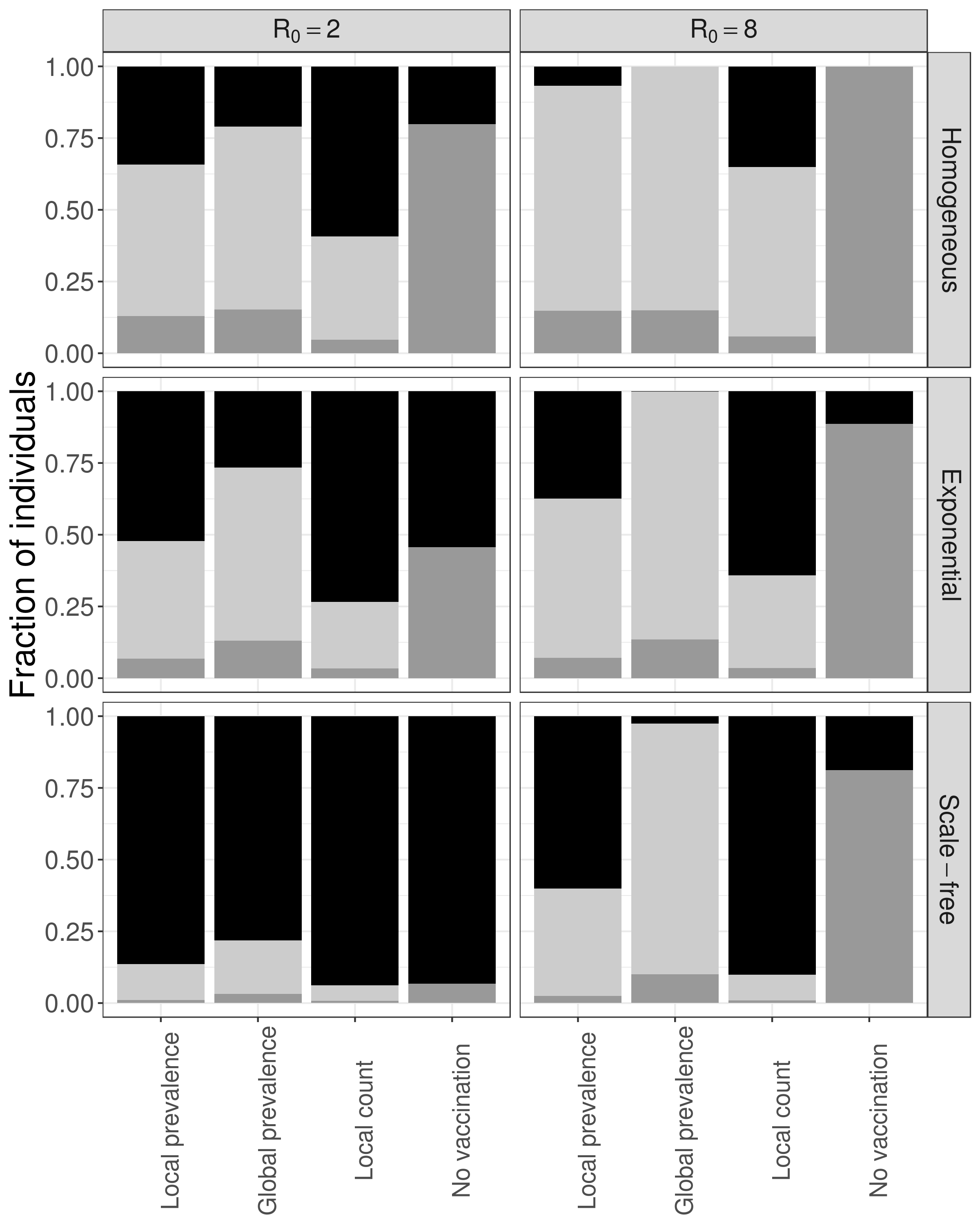}
\caption{Decision dynamics across social networks. We compare three different network structures--homogeneous (top), exponential (middle), and scale-free (bottom)--and show the susceptible (black), infected (dark gray) and vaccinated (light gray) for two different values of $R_0$. Results are averages across 500 stochastic simulations.}
\label{diffNets}
\end{figure}

Finally, we consider the impact of the underlying contact network on the interplay between transmission and vaccination dynamics (Figure \ref{diffNets}). We compare our focal exponential network to a uniform random network in which all nodes have the same degree and a Barab\'asi-Albert scale-free network. The local count strategy robustly affords the most efficient population-level protection, averting the maximum number of infections (or nearly maximum in the case of the scale-free network and low $R_0$) with the fewest vaccines.

\section{Discussion}

Public health interventions and individual-level adherence decisions can profoundly influence the fate of unfolding epidemics. In this study, we assume that individuals have access to a fully protective measure, such as self-isolation, medical prophylaxis, or an immediately and completely efficacious vaccine. We assume that individuals take action based on their perceived short-term risk of infection and compare the epidemiological impacts of three different plausible decision paradigms. They gauge personal risks based on either direct knowledge of infected friends and family (number or fraction of infected social contacts) or  indirect information about population-level prevalence, perhaps gleaned through news media. 

Of the three decision models, global risk assessments prove least effective across a large range of disease scenarios ($R_0$ ranging from one to ten). Nearly all non-infected individuals eventually vaccinate, yet the total cases more than double those occurring under the alternative strategies. There is a mismatch between risk and action. Risk is highly variable in time and space, given the heterogeneity of the underlying contact network and branching nature of transmission. Yet, the global model assumes that perceived risk and the consequent likelihood of adherence is homogeneous throughout the network, though variable in time. By the time global prevalence triggers wide-spread action, the highest risk individuals have already been exposed and the lowest risk individuals may still not, and might never, require protection.

In contrast, when individuals make decisions based on local risk assessments, the intervention efforts more closely track the epidemiological dynamics. Tallying infected contacts rather than estimating the fraction of infected contacts provides a more accurate indication of real-time risk and results in more efficient intervention. Assuming that all social contacts are equally likely to transmit disease, two out of three infected contacts carries the same immediate risk as two out of ten. The advantage of local risk assessment stems from two sources of variation in risk. First, disease transmission is an inherently local process in which risk aggregates around currently infected individuals. Second, this is magnified in realistically heterogeneous networks, by the concentration of risk around the center (most connected individuals) of the network.

Given infinite resources, all three of the decision paradigms would markedly diminish an emerging outbreak. However, interventions may be constrained by limited supplies or lack of population access to medical countermeasures, such as vaccines or antimicrobials. Even social distancing measures, such as self-isolation, may be limited by economic necessity---the need to go to work, school or daycare---or care-giving obligations for extended family. While such limitations should be formally analyzed, our simple analysis suggests that the best paradigm for averting infections also requires the fewest resources. For example, for a flu-like $R_0$ of two, compare the local count strategy, where individuals protect themselves as their number of infected friends and family climb, to the global strategy, where decisions are based on population prevalence. For every individual that takes action, almost two infections are averted under the local strategy whereas less than one infection is averted under the global strategy. Local counting results in far fewer total infections (3\% versus 13\%) while requiring far less intervention resources (23\%  versus 60\% of individuals taking protective action).

Several studies suggest that immunizing or isolating interventions should target the most connected individuals in a population \cite{pastor,goltsev,pastor1,sarsR0}. However, we rarely know the full contact network of a population. As proxies, we can target populations subgroups that tend to have high numbers of potentially disease-spreading contacts, such as young and school-aged children or health-care workers. We can also use biased sampling to identify highly connected individuals, such as the \emph{random acquaintance} strategy in which random individuals are asked to name one of their social contact; individuals with more contacts are more likely to be named \citep{newmanRipple,cohen,kitsak,christley}. In a sense, the winning paradigm of our study---counting infected contacts---similarly biases interventions efforts towards more connected parts of the network. The more connected one is, the more likely one is to have several infected contacts.  

The model is intentionally simplistic, providing a best case scenario for each of the three strategies. We assume that resources are unlimited, protection is immediate and complete, and  adherence probabilities perfectly mirror perceived risks. Furthermore, depending on the decision paradigm, individuals fairly accurately estimate the infectiousness of the disease, their number or fraction of infected social contacts, or the population average risk of infection. The model also assumes that individuals are short-sighted and make reactive decisions to avert immediate threat. We conjecture that the qualitative results of our analysis---the optimality of assessing risk based on the numbers of infected friends and family---are robust for a large class of 'on-the-fly' interventions that afford relatively rapid protection in the heat of an epidemic, but may not apply to preventative measures taken early in an outbreak or those with long efficacy lags. (For example, see alternative models presented in Supplement Figures \ref{sensitivity1} and \ref{sensitivity2}).

As a final caveat, we highlight our assumption that all edges (contacts) in our networks are equally likely to transmit disease. In reality, contacts can be highly heterogeneous, with household and health care contacts far more likely to transmit disease than casual social acquaintances. Our results should be robust when such heterogeneity is distributed randomly throughout the network. However, if individuals with more contacts tend to spend less time with each one, then epidemiological risk may be more homogeneous throughout the network and the advantage of the local decision strategies reduced. Although we do not model this scenario directly, we considered a homogeneous network where all individuals have the same number of contacts (Figure \ref{diffNets}). The local strategies still prevail, but their relative efficiency is reduced, with far more vaccines required to achieve the same benefit. 

This study prompts two practical questions. First, how do people actually make intervention decisions? Perhaps individuals fall nicely into one of these three decision-making camps. More likely, individual risk assessments are constrained by historical inertia \cite{durham,thomas,cornforth,ndeffo,taylor}, influenced by decisions of friends and family \citep{serpell,samit,mapVacc,brown,myers,ndeffo}, and integrate information from a combination of local and global data sources of variable reliability. Realistic decision models, driven by sociological survey data, can elucidate vaccine campaign failures and identify key pressure points for increasing uptake. Second, how can we streamline intervention campaigns to achieve efficient, rather than universal, adherence? This study reminds us that more intervention is not necessarily better intervention. The decision paradigm that most reduced transmission also required the least resources. Given the simplicity of our model, we do not suggest that public health agencies should promote `infection-counting'. Rather, we conclude that public health agencies should prioritize \emph{local} disease surveillance and risk communications efforts and believe that data-driven models can be instrumental in designing effective outbreak information campaigns.

\section*{Competing interests} We have no competing interests.

\section*{Author's contribution}
All authors participated in the design of the study and drafted the manuscript. All authors gave final approval for publication.

\section*{Funding} 
These project was supported by NIGMS MIDAS grant U01-GM087719. Jos\'e Luis Herrera Diestra is supported by the São Paulo Research Foundation (FAPESP) under grants 2016/01343-7 and 2017/00344-2.

\newpage
\beginsupplement
\section*{Supplemental material}

\begin{figure}[!ht]
\centering\includegraphics[width=\linewidth]{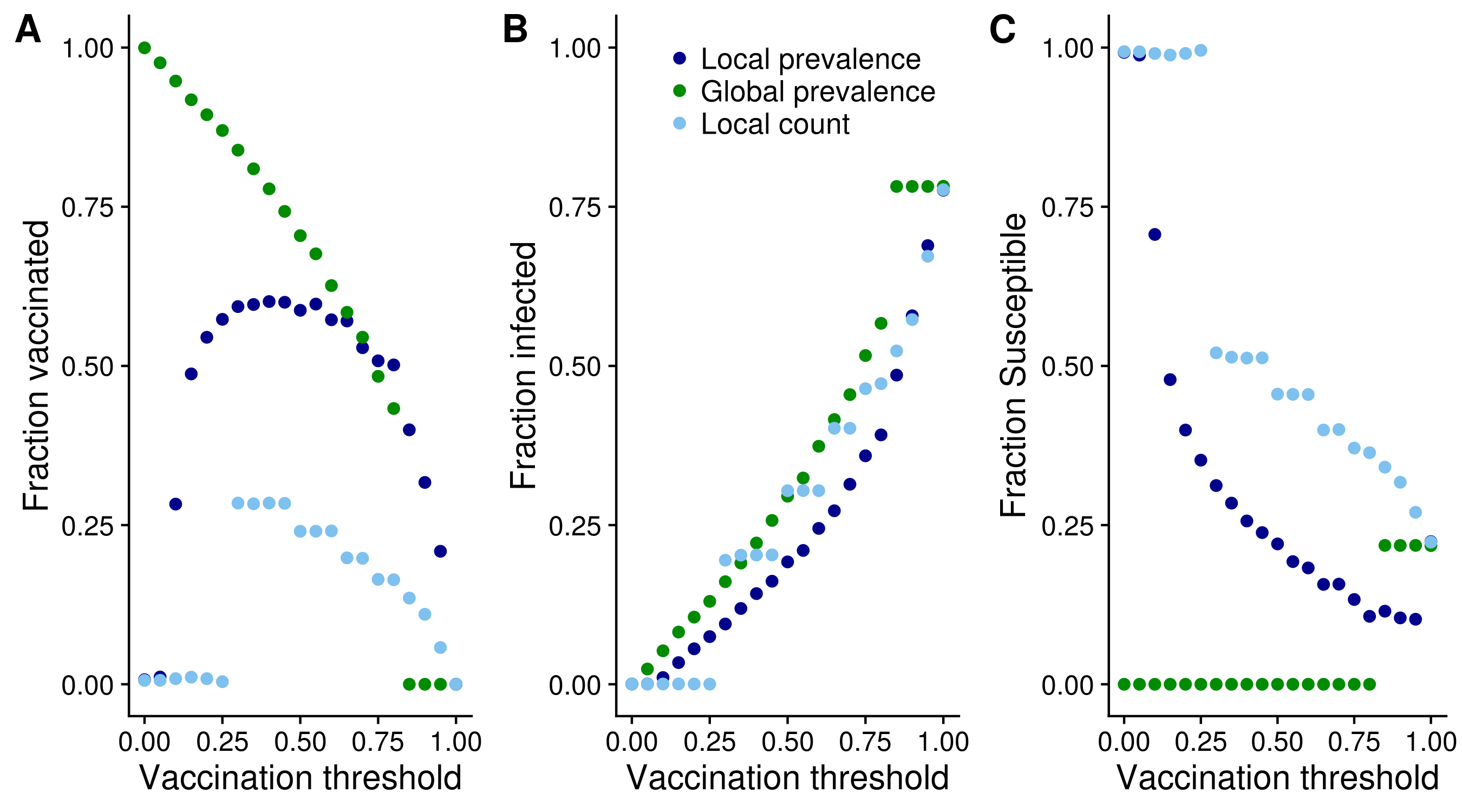}
\caption{{\bf Epidemiological outcomes for threshold-based vaccine decisions}.  Individuals determine infection risk according to local count, local prevalence or global prevalence strategies and then vaccinate when perceived risk crosses a specified vaccination threshold (given along the x-axes), rather than vaccinating probabilistically according to perceived risk. The average fraction of individuals in the population that (A) vaccinate (and are not infected), (B) become infected (with or without vaccination), and (C) remain susceptible as a function of the vaccination threshold. This analysis assumes $R_0=5$. Values are averages across 500 stochastic simulations. }
\label{sensitivity1}
\end{figure}

\begin{figure}[!ht]
\centering\includegraphics[width=0.8\linewidth]{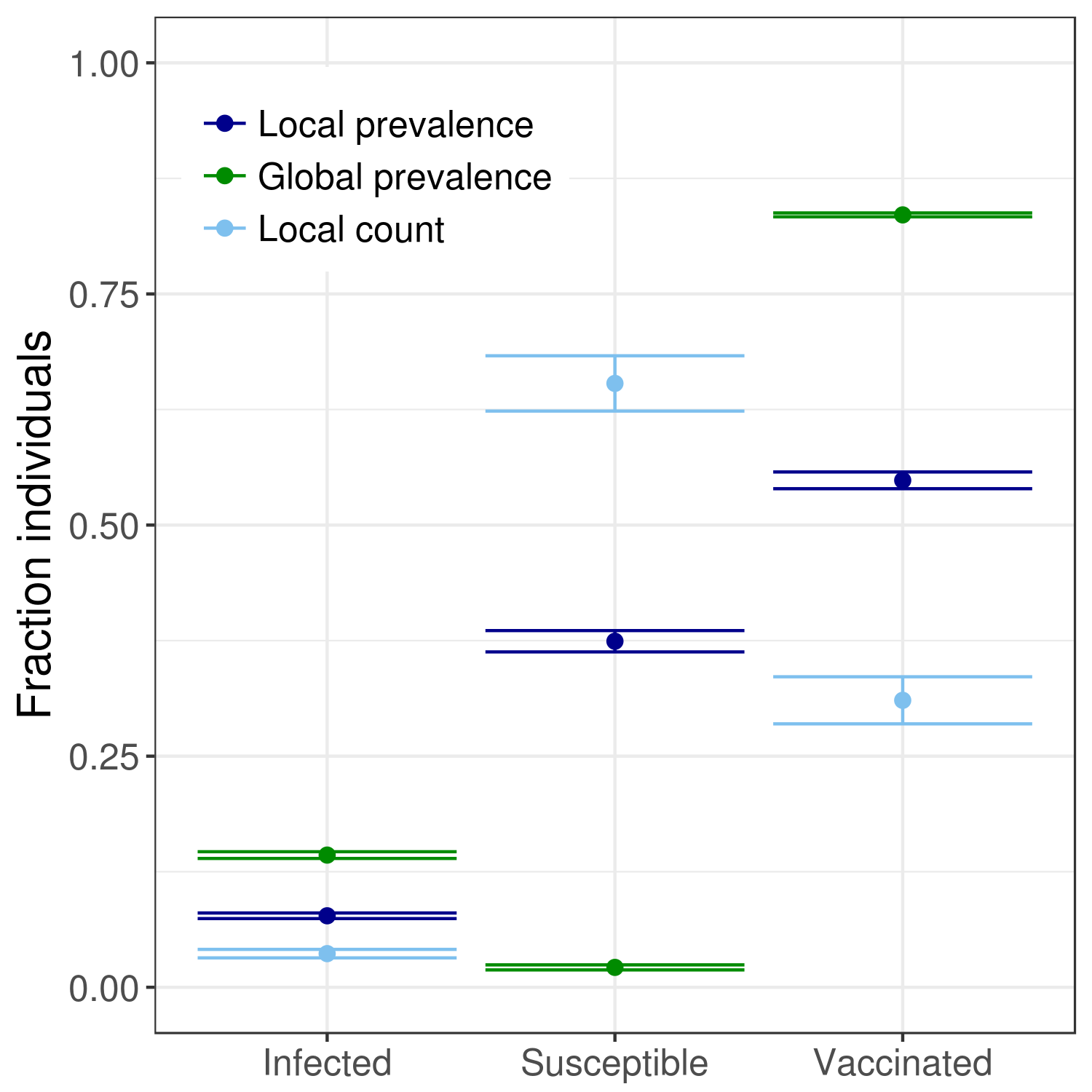}
\caption{{\bf Epidemiological outcomes for a delayed and imperfect vaccine}. Individuals follow the original vaccination strategies, but experience a one-day delay between deciding to vaccinate and becoming immune. Furthermore, the vaccine is assumed only 80\% effective, meaning that 80\% of vaccinating individuals enjoy full protection one day after deciding to vaccinate and the remaining 20\% of vaccinating individuals remain fully susceptible. The graph shows the means and standard deviations in the fraction of individuals infected across 500 stochastic simulations, assuming $R_0=5$.}
\label{sensitivity2}
\end{figure}

To evaluate the robustness of our results, we consider two alternative decision strategies. The first is a {\it threshold} model in which individuals vaccinate when their personal risk estimate (based on either the original local count, local prevalence or global prevalence models) crosses a specified threshold (Figure \ref{sensitivity1}). The second is a {\it delayed} and {\it imperfect} vaccination model in which individuals make decisions according to the original models, but vaccines are only 80\% efficacious (i.e., 80\% of vaccinators are fully protected; 20\% remain susceptible), with immunity delayed by one day after the decision to vaccinate (Figure \ref{sensitivity2}). We find qualitatively similar patterns, with the local count strategy (tracking the number of infected contacts) typically affording the greatest population-wide protection with the fewest vaccines. The exception is that the local prevalence strategy averts more infections than the local count strategy at intermediate decision thresholds, but requires approximately double the number of vaccines to do so (Figure \ref{sensitivity1}).

\end{document}